\documentclass[journal]{IEEEtran}
\usepackage{myStyleIEEE}
\usepackage{nomencl}
\usepackage{ragged2e}
\usepackage{cuted}
\usepackage[thinc]{esdiff}
\usepackage{accents}

\IEEEoverridecommandlockouts

\newlength{\bracewidth}

\newcommand{\ubar}[1]{\underaccent{\bar}{#1}}

\DeclareMathOperator{\Tr}{Tr}
\usepackage{etoolbox}
\renewcommand\nomgroup[1]{%
  \item[\bfseries
  \ifstrequal{#1}{P}{B. Parameters}{%
  \ifstrequal{#1}{V}{D. Variables}{%
  \ifstrequal{#1}{F}{A. Functions}{%
  \ifstrequal{#1}{S}{C. Sets and Indices}{}}}}%
]}

\begin{document}
% \pagecolor{cGreen}

\title{Managing the Uncertainty in System Dynamics Through Distributionally Robust Stability-Constrained Optimization}

\newtheorem{proposition}{Proposition}
\renewcommand{\theenumi}{\alph{enumi}}

\author{Zhongda~Chu,~\IEEEmembership{Member,~IEEE,} and
        Fei~Teng,~\IEEEmembership{Senior Member,~IEEE} 
        % \thanks{Zhongda Chu and Fei Teng (e-mail: f.teng@imperial.ac.uk) are with Imperial College London.
        
        % This work was supported by EPSRC under Grant EP/T021780/1.}
        % <-this % stops a space
        
\vspace{-0.5cm}}
\maketitle
\IEEEpeerreviewmaketitle

%ABSTRACT
\begin{abstract}
With the increasing penetration of Inverter-Based Resources (IBRs) and their impact on power system stability and operation, the concept of stability-constrained optimization has drawn significant attention from researchers. In order to manage the parametric uncertainty due to inaccurate modeling that influences the system dynamics, this work proposes a distributionally robust stability constraint formulation. \textcolor{black}{However, the uncertainty of system dynamic parameters influences the stability constraints indirectly through a nonlinear and implicit relationship. To address this issue, a propagation mechanism from the uncertainty of the system dynamic parameters to the stability constraint coefficients is established.} Since these coefficients are connected to the uncertain parameters through highly nonlinear and implicit functions, an approximation approach utilizing Taylor expansion and the Delta method is developed to estimate the statistical moments of the stability constraint coefficients based on the first and second-order derivatives, with which an ambiguity set for the distributionally robust optimization can be formulated. The accuracy of the uncertainty propagation as well as the effectiveness of the distributionally robust stability constraints are demonstrated through detailed case studies in the modified IEEE 39-bus system.

\end{abstract}


\makenomenclature
\renewcommand{\nomname}{\textcolor{black}{Nomenclature}}
\mbox{}
\nomenclature[V]{$\mathbf{g}$}{system stability index}
\nomenclature[V]{$X$}{decision variable in the optimization}
\nomenclature[V]{$\Tilde{\mathbf{g}}$}{approximated system stability index}
\nomenclature[V]{$\mathsf X$}{augmented decision variable}
\nomenclature[V]{$C$}{objective of regression model}
\nomenclature[V]{$\beta$}{constraints of regression model}
\nomenclature[V]{$L$}{Lagrange function of regression model}
\nomenclature[V]{$\mathsf K^*$}{optimal solution of regression model}
\nomenclature[V]{$\pmb{\lambda}$}{Lagrange function of regression model}
\nomenclature[V]{$Y_{red}$}{reduced admittance matrix}
\nomenclature[V]{$Y$}{admittance matrix}
\nomenclature[V]{$V_{\Phi(c_l)}$}{voltage at IBR $c_l$ terminal}
\nomenclature[V]{$P_{c_l}$}{Power from IBR $c_l$}
\nomenclature[V]{$X_g$}{reactance of SGs}
\nomenclature[V]{$C_g$}{operation cost of SGs}
\nomenclature[V]{$P^s$}{shed load}

\nomenclature[P]{$\mathbf{g_{\mathrm{lim}}}$}{system stability index limit}
\nomenclature[P]{$\mathsf K$}{stability constraint coefficients}
\nomenclature[P]{$\nu$}{parameter to define $\Omega_2$ and $\Omega_3$}
\nomenclature[P]{$\mathsf p$}{uncertain parameters}
\nomenclature[P]{$\mu_{\mathsf p}$}{mean of uncertain parameters}
\nomenclature[P]{$\Sigma_{\mathsf p}$}{covariance matrix of uncertain parameters}
\nomenclature[P]{$\mu_{\mathsf K}$}{mean of $\mathsf K$}
\nomenclature[P]{$\Sigma_{\mathsf K}$}{covariance matrix of $\mathsf K$}
\nomenclature[P]{$s$}{spatial scale of regression weights}
\nomenclature[P]{$\gamma$}{curve steepness of the regression constraints}
\nomenclature[P]{$\mathrm{gSCR}_\mathrm{lim}$}{limit of general short-circuit ratio}
\nomenclature[P]{$w/v$}{left/right eigenvectors of matrix $Y_{red}'$}
\nomenclature[P]{$t$}{time step in UC}
\nomenclature[P]{$c^s$}{load shedding cost}
\nomenclature[P]{$\eta$}{confidence level}
\nomenclature[P]{$\tau$}{eigenvalue of $\Sigma_{\mathsf K}$}
\nomenclature[P]{$q$}{eigenvector of $\Sigma_{\mathsf K}$}

\nomenclature[F]{$h$}{function of optimal coefficients}
\nomenclature[F]{$\mathbf{g}$}{function of stability index}
\nomenclature[F]{$f$}{function composition of $h$ and $\mathsf g$}
\nomenclature[F]{$\lambda_{\mathrm{min}}$}{minimum eigenvalue}
\nomenclature[F]{$\xi$}{probability density function}

\nomenclature[S]{$\omega\in\Omega$}{sample in data set}
\nomenclature[S]{$\Omega_{1/2/3}$}{subsets of $\Omega$}
\nomenclature[S]{$m\in\mathcal{M}_{in}$}{set of inactive constraints}
\nomenclature[S]{$m\in\mathcal{M}_{a}$}{set of active constraints}
\nomenclature[S]{$n\in\mathcal{N}$}{set of scenarios}
\nomenclature[S]{$b\in\mathcal{B}$}{set of buses}
\nomenclature[S]{$g\in\mathcal{G}$}{set of SGs}
\nomenclature[S]{$c_l\in\mathcal{C}_l$}{set of GFL}
\nomenclature[S]{$c_m\in\mathcal{C}_m$}{set of GFM}
\nomenclature[S]{$\mathcal{C}_L$}{set of GFL buses}
\nomenclature[S]{$\delta$}{set of rest buses apart from GFL buses}
\vspace{-0.7cm}
\printnomenclature

\section{Introduction}
The ongoing trend towards a clean and sustainable power system due to the environmental concerns requires wide-scale integration of Inverter-Based Resources (IBRs). Together with the retirement of conventional Synchronous Generators (SGs), these power electronics interfaced devices bring new challenges to power system operation and stability due to their significantly distinguished characteristics in power generation and conversion. Specifically, the decline of rotational inertia, frequency and voltage support as well as the loss of stable and inherent synchronization mechanism have been identified as the main threats for the future high IBR-penetrated system~\cite{dorfler2023control}. 

In order to maintain stable and secure power system operation by coordinating different resources in the system and maximize the overall economic profit, it is necessary to develop stability constraints and incorporate them during system scheduling process. The works in \cite{9925092,9066910} focus on the frequency stability issues in low-inertia systems. Voltage stability-constrained optimal system scheduling is investigated in \cite{wang2023voltage,9786660} to ensure static voltage stability by improving the power transfer capability and system strength at the IBR buses. Synchronization stability is studied in \cite{liederer2022transient,yuan2023transient} where the stability is achieved by explicitly confining the rotor angle deviations or the minimum eigenvalue of the Hessian matrix of the energy function. \textcolor{black}{On the other hand, research utilizing novel data-driven approaches to extract system security constraints in the Optimal Power Flow (OPF) problem has also been discussed in the literature \cite{halilbavsic2018data,murzakhanov2020neural}, where the dynamic security constraints are formulated based on decision trees and neural networks respectively.}

The uncertainties in the system dynamics, however, have been less focused, which may result in unstable or overconservative operation. \textcolor{black}{In order to address the uncertainty in the stability-constrained optimization, different approaches have been proposed.} The authors in \cite{xia2016probabilistic} propose probabilistic transient stability constrained optimal power flow to consider the correlated uncertain wind generation. A group search optimization and the point estimated method are further applied to solve the probabilistic-constrained optimization problem. 
% However, a single-machine equivalent method is used to derive the system transient stability constraint. 
This work is improved in \cite{su2023deep}, where a deep sigma point process is proposed to predict the transient stability model based on time-domain simulation while considering the uncertainty related to the renewable generation and the loads. 
% The resulting nonlinear programming is solved by the primal-dual interior point method. 
A coordinated method for preventive generation rescheduling and corrective load shedding to maintain power system transient stability under uncertain wind power variation is proposed in \cite{yuan2020preventive}, where the preventive and corrective control is coordinated by a risk parameter in the two-stage bi-level optimization. 
% The nonlinear stability constraints are linearized based on the extended equal area criterion and the trajectory sensitivity. 
The stability issues brought by renewable energy sources with non-Gaussian uncertainties in isolated microgrids are also investigated in \cite{wang2023stability}, where the stability chance constrained optimal power flow is formulated as a bi-level optimization with the lower level handling the stability index through semi-definite programming. 
% The Gaussian mixture model is applied to represent the non-Gaussian RES uncertainties and the analytical sensitivity analysis is used to reformulate chance constraints into linear deterministic versions.
\textcolor{black}{Note that since this work focuses on the topic of stability-constrained optimization, the reviewed research is not limited to a specific problem such as optimal power flow or Unit Commitment (UC).}

\textcolor{black}{Nevertheless, these existing works concentrate on the uncertainty of the renewable generation, which indirectly influences the system stability through the impact on operating points and the stability services, whereas the uncertainty associated with the system dynamic models due to factors such as inaccurate modeling and unknown component parameters has not been investigated.} On one hand, the parameters of the conventional system components, such as SGs, may vary due to environmental changes. On the other hand, considering the manufacturers being reluctant to share the specific control algorithms and parameters of IBRs, their accurate dynamic models are unknown to system operators, thus also suffering from uncertainty. This would become more evident in the future system with higher IBR penetration. Some work has been done to account for these uncertainties in parameter estimation and control design. A Bayesian inference framework has been proposed in \cite{xu2019adaptive} to estimate the SG parameters based on a polynomial-based reduced-order representation. The authors in \cite{amin2018gray} develop a gray-box method to estimate the parameters of wind farm controllers, based on the measurements of frequency domain equivalent impedance combined with nonparametric impedance identification. 
% The uncertainty of AC grid impedance is considered in \cite{tavakoli2022robust} for the HVDC systems during the control design process to ensure the robustness of the stability and performance. 
Reference \cite{tran2021optimized} considers the inverter model uncertainty in the active disturbance rejection control for grid-connected inverters, which guarantees a stable operation of inverters under uncertainties. However, a unified modeling and management framework to consider the uncertainties in system dynamics and their impact on system stabilities for stability-constrained optimization is yet to be investigated. In this context, this work proposes a distributionally robust stability-constrained optimization model to ensure system stability under system dynamic uncertainties. The main contributions of this paper are identified as follows:

\begin{itemize}
    % \item A stability-constrained optimization framework in a high-IBR penetrated system considering the uncertainty of the system dynamics is proposed based on the distributionally robust approach, where chance constraints on system stability are formulated in Second-Order Cone (SOC) form.
    % \item Uncertainties associated with the constraint coefficients in the optimization model as well as the stability indices due to the concerned uncertain parameters are analytically quantified, by propagating the statistical moments through nonlinear and implicit functions.
    \color{black}
    \item An uncertainty management framework for stability-constrained optimization in a high-IBR penetrated system is proposed where system dynamic parameter uncertainty within the stability index is explicitly considered by formulating a Second-Order Cone (SOC) based distributionally robust chance constraint.
    \item The uncertainty of the stability constraint coefficients due to the uncertain parameters in system dynamics are quantified by propagating the statistical moments of the uncertain parameters through a nonlinear and implicit function composition of a constrained regression model and the stability index.
    \color{black}
    \item The accuracy and the effectiveness of the proposed approach are demonstrated with the modified IEEE 39-bus system. The impact of system dynamic uncertainties and the uncertainty levels on system operation within the proposed distributionally robust stability-constrained framework is thoroughly investigated with detailed case studies.
\end{itemize}

The remainder of this paper is structured as follows. In Section~\ref{sec:2}, a unified representation of the system stability constraints is introduced. Section~\ref{sec:3} derives the uncertainty of the stability constraint coefficients from the uncertain parameters in the system dynamics. The distributionally robust stability chance-constrained system scheduling model is formulated in Section~\ref{sec:4}. An alternative formulation based on robust distributional learning is discussed in Section~\ref{sec:5}, followed by case studies in Section~\ref{sec:6}. Finally, Section~\ref{sec:7} draws the main conclusions and discusses the outlook of the study.

\section{Unified Representation of System Stability Constraints} \label{sec:2}
In order to analyze and assess system stability, a number of stability indices have been established based on the system dynamic models of different forms, such as state-space models and impedance models. Due to their close relevance to system-level features, such as system inertia, impedances, and operating points, many of these stability indices have been formulated as operational constraints during the system scheduling process, to ensure stable system operation. As a result, we express a stability index in the following general form:
\begin{equation}
    \label{g(X)}
    \mathbf{g} ({X}) \ge \mathbf{g}_{\mathrm{lim}},
\end{equation}
where ${X} \in \mathbb{R}^n$ is the decision variable in system optimization model and $\mathbf{g}_{\mathrm{lim}} \in \mathbb{R}$ is the stability index limit, above which the system remains stable. Some examples are given here to illustrate the generality of \eqref{g(X)}. Based on the system state-space model, the small-signal stability can be characterized by the eigenvalues of the closed loop state matrix $A$. Therefore, in this case, \eqref{g(X)} can be rewritten as $\Re(\lambda_{\mathrm{min}}\left(A(X))\right) < 0$, where $ X$ could represent system operating points. As for the voltage stability, many indices are proposed based on the power transfer capability \cite{hosseinzadeh2021voltage,9786660}, where the stability index can be represented as a function of the power flow injection at the concerned buses and the equivalent grid impedances. Moreover, considering the impedance model of a general multi-machine system, the small-signal stability criterion due to the grid-following (GFL) IBRs can be developed in the form of generalized short-circuit ratio \cite{gOSCR}, i.e., $\mathrm{gSCR} \ge \mathrm{gSCR}_{\mathrm{lim}}$, where the $\mathrm{gSCR}$ represents the connectivity of the network, depending on the system topology and impedances.

However, it is also understandable that, due to the strong nonlinearity, it is in general difficult or even impossible to directly include these stability indices as operational constraints in power system optimization models. A unified framework that can effectively reformulate the stability constraints, in a general way, to fit any typical power system optimization model is presented in \cite{Part_I}, which is briefly introduced below.

The target of the stability constraint reformulation is to describe the boundary of the stability feasible region through a simple structure. The estimated expression of the nonlinear functions $\mathbf{g}$ is defined as follows:
\begin{align}
    \label{g_SOC}
    \Tilde{\mathbf{g}}  (\mathsf{X}) & = \mathsf{K}^{\mathsf T} \mathsf{X}  ,
\end{align}
where $\Tilde{\mathbf{g}}\in \mathbb R$ is the estimated function of $\mathbf{g}$ and $\mathsf{K}\in\mathbb{R}^{\bar k}$ is the coefficients of $\mathsf X$. $\mathsf X\in\mathbb{R}^{\Bar{k}}$ is the augmented decision variable which may contain $1$, $X$ and products of any two elements in $X$. The resulting products of a binary variable and a continuous terms or bilinear terms can be dealt with through Big-M and binary expansion, which is not discussed here. 

Substituting \eqref{g_SOC} into \eqref{g(X)} leads to the reformulated system stability constraint as:
\begin{align}
    \label{linear_glim}
    \mathsf{K}^{\mathsf T} \mathsf{X} \ge \mathbf{g}_{\mathrm{lim}}.
\end{align}
In order to find the coefficients $\mathsf{K}$ that best describes the boundary of the system stability feasible region, the following boundary-aware optimization is utilized:
\begin{subequations}
\label{DM3}
\begin{align}
    \label{obj3}
    \min_{\mathsf{K}}\quad & \sum_{\omega \in \Omega_2} \left(\mathbf{g}^{(\omega)} - \Tilde{\mathbf{g}}^{(\omega)} \right)^2\\
    \label{coef_ctr2}
    \mathrm{s.t.}\quad & \Tilde{\mathbf{g}}^{(\omega)}< \mathbf{g}_{{\mathrm{lim}}},\,\,\forall \omega \in \Omega_1\\
    \label{coef_ctr3}
    &\Tilde{\mathbf{g}}^{(\omega)} \ge {\mathbf{g}}_{{\mathrm{lim}}},\,\,\forall \omega \in \Omega_3,
\end{align}
\end{subequations}
with $\omega = \left(\mathsf{X}^{(\omega)},\, \mathbf{g}^{(\omega)}\right)\in \Omega$ being the entire data set corresponding to the stability constraint and $(\cdot)^{(\omega)}$ denoting the quantity evaluated at data point $\omega$. The data set $\Omega$ is generated by evaluating $\mathbf{g}$ in representative system conditions. The sets $\Omega_1 ,\, \Omega_2$ and $\Omega_3 $ are the subsets of $\Omega$, whose relationship is defined as below:
\begin{subequations}\label{Omega}
\begin{align}
    \Omega &= \Omega_1 \cup\Omega_2\cup\Omega_3 \\
    \label{Omega1}
    \Omega_1 & = \left\{\omega\in \Omega \mid \mathbf{g}^{(\omega)}<\mathbf{g}_{{\mathrm{lim}}} \right\}\\
    \label{Omega2}
    \Omega_2 & = \left\{\omega\in \Omega \mid \mathbf{g}_{{\mathrm{lim}}} \le \mathbf{g}^{(\omega)}<\mathbf{g}_{{\mathrm{lim}}} + \nu \right\}\\
    \label{Omega3}
    \Omega_3 & = \left\{\omega\in \Omega \mid \mathbf{g}_{{\mathrm{lim}}} + \nu\le \mathbf{g}^{(\omega)} \right\},
\end{align}
\end{subequations}
with $\nu$ being a constant parameter. Given \eqref{coef_ctr2} and \eqref{Omega1}, all the data points whose real stability indices, $\mathbf{g}^{(\omega)}$ are smaller than the limits, can be identified correctly by the estimated function, $\Tilde{\mathbf{g}}^{(\omega)}$. Ideally, it is also desired to correctly identify all the above-limit data points, which would make it a classification problem. However, this may cause infeasibility due to the restricted structure defined in \eqref{g_SOC}. Therefore, a parameter $\nu\in \mathbb{R}^+$ is introduced to define $\Omega_2$ and $\Omega_3$ as in \eqref{Omega2} and \eqref{Omega3}. In this way, all the data points in $\Omega_3$ will be classified correctly and misclassification can only occur in $\Omega_2$, thus being conservative. Furthermore, $\nu$ should be chosen as small as possible while ensuring the feasibility of \eqref{DM3}.

\section{Uncertainty Quantification} \label{sec:3}
This work focuses on the uncertainty associated with the system dynamic model, such as the impedances of SGs and IBRs. These parameters influence the system stability performance and thus the stability constraints, which eventually bring uncertainty to the constraint coefficients $\mathsf K$. In this section, the uncertainty of the stability constraint coefficients is derived analytically by propagating the statistical moments through nonlinear and implicit functions from uncertain system parameters.

\subsection{Uncertainty Representation} \label{sec:3.1}
Although the parameters of different system components such as generators, transformers and transmission lines are typically provided by manufacturers, their actual values can deviate substantially from manufacturer ones \cite{9020274}, due to a number of factors including magnetic saturation, internal and ambient temperature, unit aging, and the effect of centrifugal forces on winding contacts and incipient faults within the machines \cite{karayaka2003synchronous}. \textcolor{black}{Moreover, since the detailed control logic and parameters of the commercial IBRs are typically not shared by the manufacturers, their exact dynamic performance cannot be modeled perfectly by system operations \cite{li2022systematic}.} Methodologies have also been proposed for model identifications and parameter estimations of convectional SGs and IBRs based on either offline testing methods or online measuring methods \cite{xu2019adaptive,song2020parameter}. However, good accuracy cannot be guaranteed due to measurement errors and impractical assumptions \cite{xu2019adaptive,amin2018gray}, which has to be accounted for in the stability-constrained optimization.

We first redefine the nonlinear expression of the stability index $\mathbf g$ as a function of the decision variables $\mathsf X$ and the uncertain parameters $\mathsf p$, which converts \eqref{g(X)} into:
\begin{align}
    \label{gi}
    \mathbf{g}(\mathsf{X}, \mathsf p) \ge \mathbf{g}_{\mathrm{lim}}.
\end{align}
Based on the prior knowledge about the uncertainty, different approaches can be applied to manage the uncertainty in the stability-constrained optimization. For instance, robust optimization can be formulated if only the bounds of uncertain parameters are known:
\begin{subequations}
\begin{align}
    \min_{\mathsf X\in \mathcal{X}} \quad & J(\mathsf X) \\
    \mathrm{s.t.} \quad & \min_{\mathsf p\in[\ubar{\mathsf p}, \bar{\mathsf p}]} \mathbf{g}(\mathsf{X}, \mathsf p) \ge \mathbf{g}_{\mathrm{lim}},
\end{align}    
\end{subequations}
where $J(\mathsf X)$ represents the objective function of the stability-constrained optimization, (e.g. system operation cost), $\mathcal{X}$ the feasible region determined by other constraints and $\mathsf p\in[\ubar{\mathsf p}, \bar{\mathsf p}]$ the uncertain set. Alternatively, if a detailed distribution of the uncertainty is available, a chance-constrained optimization in the following form can be developed:
\begin{subequations}
\begin{align}
    \min_{\mathsf X\in \mathcal{X}} \quad & J(\mathsf X) \\
    \mathrm{s.t.} \quad & \mathrm{Pr} \left\{\mathbf{g}(\mathsf{X}, \mathsf p) \ge \mathbf{g}_{\mathrm{lim}} \right\} \ge \eta,
\end{align}    
\end{subequations}
with $\eta$ being the pre-defined confidence level. 

% \textcolor{black}{However, the bounds or the detailed distribution of the uncertain parameters may not be available to system operators, and simply assume ``reasonable'' values for the bounds or Gaussian distribution may be unrealistic and lead to biased estimation \cite{xu2019adaptive}. Therefore, this work focuses on a distributionally robust formulation to deal with the situation where only the statistical moments are known based on the information from manufacturers or historical data and measurements, whereas the specific probabilistic distribution is unknown.} 

The general expression takes the form:
\begin{subequations}
\begin{align}
    \min_{\mathsf X\in \mathcal{X}} \quad & J(\mathsf X) \\
    \mathrm{s.t.} \quad & \min_{\mathbf{D}\in\mathcal{P}} \mathrm{Pr} \left\{\mathbf{g}(\mathsf{X}, \mathsf p) \ge \mathbf{g}_{\mathrm{lim}} \right\} \ge \eta, \label{DRO_0}
\end{align}    
\end{subequations}
where $\mathbf{D}$ is the uncertain parameter distribution and $\mathcal{P}$ the ambiguity set. \textcolor{black}{In this work, we consider the ambiguity set based on the first- and second-order moments, which is widely used for the distributionally robust formulation in the literature \cite{9475967}. However, most of the existing work considers linear constraint in optimization \cite{9475967} or LTI system with linear decision rules or affine feedback policies in control design \cite{van2015distributionally}. As for our concerned stability-constrained optimization problem, the stability index is nonlinear in terms of both decisions and uncertain variables. How to manage this type of problem in optimization has not been investigated to the best of the authors' knowledge.}

\textcolor{black}{In the concerned problem, the uncertainties come from the impedance of the system components, such as the generation sources (SGs and IBRs). The key parameters of SGs are provided by the manufacturers, which can typically represent their characteristics during normal operation. Hence it is common (and indeed recommended) practice to choose a value equal to or close to the manufacturer-supplied values \cite{SG_guide}. From this aspect, it is reasonable to assume these values can represent the means. On the other hand, if the manufacturer data is not available or known to be untrustable, different parameter estimation approaches have also been proposed in the literature, many of which can provide (asymptotically) unbiased estimation, e.g., \cite{zhou2013estimation}. As for the IBRs whose detailed dynamic models are typically not shared by manufacturers, the identification of their dynamic parameters has also drawn the attention of the researchers by utilizing data-driven approaches \cite{fan2020time}, based on which the expected value and variation range/confidence level of the dynamic parameters can be extracted \cite{jordan2015machine}. Moreover, for practical application, although the assertion that the moments are exactly known is never true, we can always ensure the approach to be conservative by slightly increasing the variance, the sensitivity analysis of which is demonstrated in Section~\ref{sec:6.3}.\\
\indent In the case that a mean value is not available with confidence, there are other distributionally robust optimization methods based on distances or likelihoods that can be potentially explored. However, how to propagate the uncertainty information under these circumstances while considering the two layers of nonlinear mappings in the stability-constrained optimization requires further investigation.}

 % \textcolor{black}{Furthermore, the uncertainty of dynamic parameters can be divided into two layers. The first layer is the parameter uncertainty characteristics could be different and uncertain at different operating conditions whilst the second is the uncertainty of the dynamic parameters at a specific operating condition due to the imperfectly known system model. However, it is unclear how the operating conditions would influence the uncertainty characteristics of the source impedance and very limited research can be found in this area in the literature. As a result, this work instead combines these two layers of uncertainties into one, and the dependence of the parameter uncertainty on the operating conditions is accounted for in its variance. If more information about this decision dependence can be revealed in the future, we will extend this work to include that dependence.}

\subsection{Uncertainty Propagation} \label{sec:3.2}
Due to the complex relationship between the stability index $\mathbf{g}$ and $\mathsf X$, $\mathsf p$, it is challenging enough to reformulate the original nonlinear constraint $\mathbf{g}(\mathsf{X}, \mathsf p) \ge \mathbf{g}_{\mathrm{lim}}$ into a mathematically tractable form, let alone its distributionally robust form \eqref{DRO_0}. Therefore, we consider a simplified representation as in \eqref{linear_glim}. Derived based on \eqref{DM3} and \eqref{Omega}, this reformulated stability constraint transforms the uncertainty from system parameters $\mathsf p = [\mathsf p_1,...,\mathsf p_p,...,\mathsf p_{\Bar{p}}]^{\mathsf T}\in \mathbb R^{\Bar{p}}$ with $\Bar{p}$ being the total number of uncertain parameters, to the stability constraint coefficients, $\mathsf K = [\mathsf K_1,...,\mathsf K_k,...,\mathsf K_{\Bar{k}}]^{\mathsf T} \in \mathbb R^{\Bar{k}}$, with $\Bar{k}$ being the total number of the coefficients.

However, it is challenging to identify the uncertainty associated with these coefficients, due to their complex dependence on the uncertain parameters. In order to achieve this, we define $\mathsf K_k$ as the following implicit function of the training data, $\left(\mathsf{X}^{\Omega}, \mathbf{g}^{\Omega}\right)$:
\begin{align}
    \label{h(X,g)}
    \mathsf K_k & = h_k\left(\mathsf{X}^{\Omega}, \mathbf{g}^{\Omega}\left(\mathsf{X}^{\Omega}, \mathsf p\right) \right) ,
\end{align}
since it is determined by solving the optimization problem \eqref{DM3}. In \eqref{h(X,g)}, $\mathsf{X}^{\Omega} = \left[...,\mathsf{X}^{(\omega)},...\right]^{\mathsf{T}}$ is the vector containing $\mathsf{X}^{(\omega)},\,\forall \omega\in \Omega$ and similarly, $\mathbf{g}^{\Omega} (\mathsf{X}^{\Omega}, \mathsf p) = \left[...,\mathbf{g}^{(\omega)},... \right]= \left[...,\mathbf{g}(\mathsf{X}^{(\omega)},\mathsf p),...\right]$. Since it is the uncertainty of the system parameter $\mathsf p$ that is of interest here, the decision variable without uncertainty $\mathsf{X}^{\Omega}$ in \eqref{h(X,g)} is omitted in the following derivation for clarity, i.e., $\mathsf K_k = h_k\left(\mathbf{g}^{\Omega}\left( \mathsf p\right) \right) = h_k \circ \mathbf{g}^{\Omega}(\mathsf p) \dot=  f_k(\mathsf p)$. Next, we derive the expectation and variance of $\mathsf K_k$ from that of $\mathsf p$. For the expectation, Taylor expansion is utilized. Expand $ f_k(\mathsf p)$ in a Taylor series around the expectation of uncertain parameters $\mu_{\mathsf p}$:
\begin{align}
\label{taylor}
    f_k(\mathsf p) & =f_k(\mu_{\mathsf p}) + \nabla {f_k} (\mu_{\mathsf p}) (\mathsf p-\mu_{\mathsf p}) \nonumber \\
    & + \frac{1}{2} (\mathsf p-\mu_{\mathsf p})^{\mathsf{T}} H_{f_k}(\mu_{\mathsf p}) (\mathsf p-\mu_{\mathsf p}) + \mathcal{O}\left((\mathsf p-\mu_{\mathsf p})^3\right),
\end{align}
where $\nabla {f_k}$ and $H_{f_k}$ are the gradient vector and Hessian matrix of $f_k$ with respect to the uncertain parameter $\mathsf p$. Taking the expectation of both sides of \eqref{taylor} and neglecting the higher order term give:
\begin{align}
\label{E_fk}
    \mathbb{E}(f_k(\mathsf p)) & \approx f_k(\mu_{\mathsf p}) + \nabla {f_k} (\mu_{\mathsf p})\mathbb E  (\mathsf p-\mu_{\mathsf p}) + \Tr[H_{f_k}(\mu_{\mathsf p}) \Sigma_{\mathsf p}] \nonumber \\
    & +\frac{1}{2} \mathbb E (\mathsf p-\mu_{\mathsf p})^{\mathsf{T}} H_{f_k}(\mu_{\mathsf p}) \mathbb E (\mathsf p-\mu_{\mathsf p})  \nonumber \\
    & = f_k(\mu_{\mathsf p}) + \Tr[H_{f_k}(\mu_{\mathsf p}) \Sigma_{\mathsf p}],
\end{align}
where $\Sigma_{\mathsf p}$ is the covariance matrix of $\mathsf p$ and the equality holds since $\mathbb E (\mathsf p - \mu_{\mathsf p}) = \mathbb E (\mathsf p) - \mu_{\mathsf p} = 0$. Further assuming that different uncertain parameters in $\mathsf p$ are independent with each other, simplifies \eqref{E_fk} as follows:
\begin{equation}
\label{E_K}
    \mu_{\mathsf K_k} = \mathbb{E}(f_k(\mathsf p)) \approx f_k(\mu_{\mathsf p}) + \sum_{p=1}^{\Bar{p}} \frac{\partial^2 f_k}{\partial \mathsf p_p^2}(\mu_{\mathsf p}) \sigma_{\mathsf p_p}^2,
\end{equation}
with $\sigma_{\mathsf p_p}^2$ being the variance of $\mathsf p_p$. \textcolor{black}{Note that it is the uncertainty of source (IBRs and SGs) impedance that is considered in this work. Since those units operate independently from each other, i.e., one generator's condition has little impact on those of others, the correlation between them can be neglected. Nevertheless, if there are indeed correlations between some parameter uncertainties, the formulation in (12) can still be utilized to capture such correlation.}

As for the covariance of $\mathsf K = [\mathsf K_1,...,\mathsf K_k,...,\mathsf K_{\Bar{k}}]^{\mathsf T} = [f_1(\mathsf p),...,f_k(\mathsf p),...,f_{\Bar{k}}(\mathsf p)]^{\mathsf T}=f(\mathsf p)$, it can be derived by applying the Delta method \cite{benichou1989delta}, which gives the following results:
\begin{equation}
\label{Cov_K}
    \Sigma_{\mathsf K} = \mathrm{Cov} (f(\mathsf p)) \approx \nabla f (\mu_{\mathsf p}) \Sigma_{\mathsf p} \nabla f (\mu_{\mathsf p}) ^{\mathsf T},
\end{equation}
where $\nabla f \in \mathbb R^{\Bar{k}\times\Bar{p}}$ denotes the Jacobian matrix of $f$ with $\frac{\partial f_k}{\partial \mathsf p_p}$ being the element in the $k$-th row and $p$-th column. \textcolor{black}{Note that the above moments propagation methods are derived based on Taylor expansion of functions of random variables and hence the performance is good only if Taylor expansion provides a good approximation of the function. The relationship between $f$ and $\mathsf p$ and the overall performance of the proposed method are also assessed in Section~\ref{sec:6.1}.} Based on \eqref{E_K} and \eqref{Cov_K}, the first and second-order moment of $\mathsf K$ can be estimated with the second and first-order derivative of $f (\mathsf p)$, if the mean and variance of $\mathsf p$ are assumed to be known. However, with $f (\mathsf p) = (h \circ \mathbf{g}^{\Omega}) (\mathsf p) $ being the composition of $h (\mathbf{g}^{\Omega})$ and $\mathbf{g}^{\Omega}(\mathsf p)$, these derivatives ($\nabla f$) cannot be determined directly. Apply the chain rule to $\frac{\partial f_k}{\partial \mathsf p_p}$ leading to:
\begin{equation}
\label{chain_1}
    \frac{\partial f_k}{\partial \mathsf p_p} = \frac{\partial h_k}{\partial \mathbf{g}^{\Omega}} \cdot \frac{\partial \mathbf{g}^{\Omega}}{\partial \mathsf p_p},
\end{equation}
with $\frac{\partial h_k}{\partial \mathbf{g}^{\Omega}} \in \mathbb R^{1\times |\Omega|}$ and $\frac{\partial \mathbf{g}^{\Omega}}{\partial \mathsf p_p} \in \mathbb R^{|\Omega|\times 1}$ being the partial derivative of coefficient $\mathsf K_k$ with respect to the stability index $\mathbf{g}^{\Omega}$ and the the stability index $\mathbf{g}^{\Omega}$ with respect to the uncertain parameter $\mathsf p_p$ respectively, which are derived in the following sections.

\subsection{Perturbation Analysis of the Regression Model} \label{sec:3.3}
In order to derive $\frac{\partial h_k}{\partial \mathbf{g}^{\Omega}}$, which is implicitly defined through the optimization problem as in \eqref{DM3}, the perturbation analysis of the regression model is carried out. Understandably, for $\frac{\partial h_k}{\partial \mathbf{g}^{\Omega}}$ to be well-defined, the optimal solution to \eqref{DM3} has to be continuous and differentiable  with respect to $\mathbf{g}^{\Omega}$. However, when $\mathbf{g}^{\omega}$ changes, due to the definition in \eqref{Omega}, the region to which the data point $\omega$ belong may change, leading to different constraints and/or objective functions and potentially discontinuous optimal solution variation. To solve this issue, the discontinuous region division in \eqref{Omega} is embedded into the optimization model \eqref{DM3}. 

For the objective function \eqref{obj3}, since only the data points in $\Omega_2$ are of concern, without loss of generality, Gaussian weights are applied to all the data points in order to increase the weights of those in $\Omega_2$ and decrease the weights of the rest:
\begin{equation}
    \label{obj3_cont}
     \sum_{\omega \in \Omega} \left(\mathbf{g}^{(\omega)} - \Tilde{\mathbf{g}}^{(\omega)} \right)^2 \exp{\left(-\frac{\left(\mathbf{g}^{(\omega)}-\left(\mathbf{g}_{\mathrm{lim}}+\frac{\nu}{2}\right)\right)^2}{2s^2}\right)},
\end{equation}
with $\Tilde{\mathbf{g}}^{(\omega)} = \mathsf{K}^{\mathsf T} \mathsf{X}^{(\omega)}$ being the linearized stability index evaluated at $\omega$ and $s$ governing the spatial scale of the weight function. Note that the Gaussian weights are not required to have a probabilistic interpretation in this application. In addition, \eqref{coef_ctr2} and \eqref{coef_ctr3} are modified as soft constraints:
\begin{subequations}
\label{coef_ctr_cont}
\begin{align}
    \label{coef_ctr2_cont}
    \Tilde{\mathbf{g}}^{(\omega)} - \gamma(\mathbf{g}^{(\omega)}-\mathbf{g}_{\mathrm{lim}}) M & \le \mathbf{g}_{\mathrm{lim}}, & &\forall \omega \in \Omega\\
    \label{coef_ctr3_cont}
    \Tilde{\mathbf{g}}^{(\omega)} + \gamma(\mathbf{g}_{\mathrm{lim}}+\nu-\mathbf{g}^{(\omega)}) M & \ge \mathbf{g}_{\mathrm{lim}}, & &\forall\omega\in \Omega
\end{align}
\end{subequations}
where $M$ is a large enough constant and the function $\gamma(\cdot)$ takes the form:
\begin{equation}
    \gamma(x) = \frac{1}{1+e^{-2rx}},
\end{equation}
with the constant $r$ being the curve steepness. 
\color{black}
Regarding the parameter tuning, $s$ and $r$ are introduced to transform the discontinuous region division into a continuous form. For $s$, which is similar to the standard deviation of the normal distribution, it is tuned such that larger weights are given to the samples in $\Omega_2$ and lower weights to the rest. Without loss of generality, the weight of 0.5 is assigned to the boundary of $\Omega_2$, i.e., $\mathbf{g}^{(\omega)}$ = $\mathbf{g}_{\mathrm{lim}}$ and $\mathbf{g}^{(\omega)}$ = $\mathbf{g}_{\mathrm{lim}} + \nu$:
    \begin{equation*}
        \exp{\left(-\frac{\left(\mathbf{g}^{(\omega)}-\left(\mathbf{g}_{\mathrm{lim}}+\frac{\nu}{2}\right)\right)^2}{2s^2}\right)} = 0.5.
    \end{equation*} 
Hence, the value of $s$ can be determined after the selection of $\nu$. As for the parameter $r$, it is the steepness of the sigmoid curve, which is applied to distinguish the boundary between the region $\Omega_2$ and $\Omega_1$/$\Omega_3$. Since a small steepness may lead to a bad performance in distinguishing the regions while a large steepness increases the ``discontinuity" of the regression model, the normalized function is selected here such that the slope of $\gamma(x)$ evaluated at $x=0$ is $1$, i.e., $r=0.5$. 
\color{black}

It should be noted that given the above reformulation, the objective function and constraints in the optimization apply to all the data points thus eliminating the strict region division in \eqref{Omega}. Rewrite the optimization model defined in \eqref{obj3_cont} and \eqref{coef_ctr_cont} in a compact form:
\begin{subequations}
\label{DM_compact}
    \begin{align}    
    \min_{\mathsf K} \quad & C(\mathsf K, \mathbf{g}^{\Omega}) \\
    \mathrm{s.t.} \quad & \beta (\mathsf K, \mathbf{g}^{\Omega}) \ge 0.
    \end{align}    
\end{subequations}
As a result, $h$ can be defined as
$ h(\mathbf{g}^{\Omega})= \arg \min_{\mathsf K : \beta(\mathsf K, \mathbf{g}^{\Omega})\ge 0} C(\mathsf K, \mathbf{g}^{\Omega})$. The Lagrange function of the optimization problem \eqref{DM_compact} at the solution point $\mathsf K^*$ is then written as:
\begin{equation}
    L(\mathsf K, \pmb{\lambda}) = C(\mathsf K, \mathbf{g}^{\Omega}) - \pmb{\lambda}^{\mathsf T} \beta (\mathsf K, \mathbf{g}^{\Omega}),
\end{equation}
with $\pmb{\lambda}$ being the Lagrangian multiplier. The Karush–Kuhn–Tucker (KKT) conditions of \eqref{DM_compact} take the form:
\begin{subequations}
\label{KKT}
\begin{align}
    \frac{\partial C (\mathsf K, \mathbf{g}^{\Omega})}{\partial \mathsf K_k}-\sum_m {\lambda_m} \frac{\partial \beta_m(\mathsf K)}{\partial \mathsf K_k} &= 0, & &k = 1,2, ..., \Bar{k} \\
    {\lambda_m} \beta_m (\mathsf K) &= 0, & &\forall m \in \mathcal{M}\\
    {\lambda_m} &\ge {0}, & &\forall m \in \mathcal{M},
\end{align}    
\end{subequations}
where $m\in \mathcal{M}$ is the set of constraints. The sensitivity of the optimal solution $\mathsf K_k$ with respect to the parameter $\mathbf{g}^{(\omega)}$ can be found from perturbation analysis of the KKT conditions. Applying a small perturbation $d\mathbf{g}^{(\omega)}$ to \eqref{KKT} gives:
\begin{subequations}
\label{KKT_perturb}
\begin{align}
    \frac{\partial^2 C }{\partial \mathsf K_k \partial \mathbf{g}^{(\omega)}}  + \sum_{k'=1}^{\Bar{k}} \frac{\partial^2 C }{\partial \mathsf K_k \partial \mathsf K_{k'}} \frac{d \mathsf K_{k'}}{d\mathbf{g}^{(\omega)}} 
     -\sum_{m\in\mathcal{M}} \left[ \frac{\partial \beta_m}{\partial \mathsf K_k} \frac{d \lambda_m}{d\mathbf{g}^{(\omega)}} \vphantom{\sum_{k'=1}^{\Bar{k}}} \right. &  \nonumber \\
    + \lambda_m \left. \left(\frac{\partial^2 \beta_m}{\partial \mathsf K_k \partial \mathbf{g}^{(\omega)}} +  \sum_{k'=1}^{\Bar{k}} \frac{\partial^2 \beta_m }{\partial \mathsf K_k \partial \mathsf K_{k'}} \frac{d \mathsf K_{k'}}{d\mathbf{g}^{(\omega)}}   \right) \right] = 0&,\, \forall k \\
    \beta_m \frac{d \lambda_m}{d\mathbf{g}^{(\omega)}} + \lambda_m \left( \frac{\partial \beta_m}{\partial \mathbf{g}^{(\omega)} }  +  \sum_{k'=1}^{\Bar{k}} \frac{\partial \beta_m }{\partial \mathsf K_{k'}} \frac{d \mathsf K_{k'}}{d\mathbf{g}^{(\omega)}} \right)  =  0 & ,\, \forall m
\end{align}    
\end{subequations}
Equation \eqref{KKT_perturb} can be further simplified based on the following facts:
\begin{subequations}
\label{KKT2}
\begin{align}
    \lambda_m & = 0, & & \forall m \in \mathcal{M}_{\mathrm{in}}\\
    \frac{d \lambda_m}{d\mathbf{g}^{(\omega)}} & =0, & &\forall m \in \mathcal{M}_{\mathrm{in}} \label{dlambda_dg} \\
    \lambda_m & \neq 0, & & \forall m \in \mathcal{M}_{\mathrm{a}}\\
    \beta_m & =0, & &\forall m \in \mathcal{M}_{\mathrm{a}} \\
    \frac{\partial \beta_m }{\partial\mathbf{g}^{(\omega)}} & = 0, & &\forall m \in \mathcal{M}_{\mathrm{a}} \label{dbeta_dg}
\end{align}
\end{subequations}
where $\mathcal{M}_{\mathrm{in}}$ and $\mathcal{M}_{\mathrm{a}}$ represent the set of inactive and active constraints. Equations \eqref{dlambda_dg} and \eqref{dbeta_dg} hold as it is assumed that the sets of active and inactive constraints remain the same after the perturbation $d\mathbf{g}^{(\omega)}$, which is an appropriate assumption since the analysis is local and the derivatives of the objective function are continuous at the optimum \cite{enevoldsen1994sensitivity,CARO20111071}. 

Combining \eqref{KKT_perturb} and \eqref{KKT2} gives the following:
\begin{equation}
\label{d_KKT_matrix}
    \begin{bmatrix}
        \mathbf{A}_{\Bar{k}\times\Bar{k}} & \mathbf{B}_{\Bar{k}\times|\mathcal{M}_a|} \\
        \mathbf{B}^{\mathsf T}_{|\mathcal{M}_a| \times \Bar{k}} & \mathbf{0}_{|\mathcal{M}_a| \times|\mathcal{M}_a|}
    \end{bmatrix} \begin{bmatrix}
        \frac{d \mathsf K}{d\mathbf{g}^{(\omega)}} \\
        \frac{d \pmb{\lambda}_{|\mathcal{M}_a|}}{d\mathbf{g}^{(\omega)}}
    \end{bmatrix} = \begin{bmatrix}
        -\mathbf{c}_{\Bar{k}}\\
        \mathbf{d}_{|\mathcal{M}_a|}
    \end{bmatrix}.
\end{equation}
$\mathbf{0}$ is the zero matrix and $\mathbf{A},\,\mathbf{B},\,\mathbf{c}$ and $\mathbf{d}$ are matrices and vectors with comfortable dimensions, in which the elements are  defined as follows:
\begin{subequations}
\label{ABcd}
    \begin{align}
        \mathbf{A}_{kk'} & = \frac{\partial^2 C }{\partial \mathsf K_k \partial \mathsf K_{k'}} -\sum_{m\in\mathcal{M}_a} \lambda_m \frac{\partial^2 \beta_m }{\partial \mathsf K_k \partial \mathsf K_{k'}}\\
        \mathbf{B}_{km} & = \frac{\partial \beta_m}{\partial \mathsf K_k}\\
        \frac{d \mathsf K}{d\mathbf{g}^{(\omega)}} & = \begin{bmatrix}
            \frac{d \mathsf K_1}{d\mathbf{g}^{(\omega)}} & ... & \frac{d \mathsf K_{\Bar{k}}}{d\mathbf{g}^{(\omega)}}
        \end{bmatrix} ^{\mathsf T} \\
        \frac{d \pmb{\lambda}_{|\mathcal{M}_a|}}{d\mathbf{g}^{(\omega)}} & = \begin{bmatrix}
            \frac{d \lambda_1}{d\mathbf{g}^{(\omega)}} & ... & \frac{d \lambda_{|\mathcal{M}_a|}}{d\mathbf{g}^{(\omega)}}
        \end{bmatrix} ^{\mathsf T} \\
        \mathbf{c}_k & = \frac{\partial^2 C }{\partial \mathsf K_k \partial \mathbf{g}^{(\omega)}} - \sum_{m\in\mathcal{M}_a} \lambda_m \frac{\partial^2 \beta_m }{\partial \mathsf K_k \partial \mathbf{g}^{(\omega)}}\\
        \mathbf{d}_m & = \frac{\partial \beta_m}{\partial \mathbf{g}^{(\omega)}}.
    \end{align}
\end{subequations}
Given the linear system \eqref{d_KKT_matrix}, the sensitivity of the optimal solution ($\mathsf K_k$) with respect to the optimization parameters ($\mathbf g^{\Omega}$) can be solved by evaluating the first- and second-order partial derivatives in \eqref{ABcd}.

The closed form expression of the first- and second-order partial derivatives in \eqref{ABcd} can be derived by combining \eqref{g_SOC} and \eqref{DM_compact}:
\begin{subequations}
\label{1st&2nd_derivatives}
\begin{align}
    \frac{\partial^2 C }{\partial \mathsf K_k \partial \mathsf K_{k'}} & = \sum_{\omega\in\Omega} \left( 2 \mathsf{X}_k^{(\omega)} \mathsf{X}_{k'}^{(\omega)} \mathsf e^{(\omega)} \right) \\
    \frac{\partial^2 \beta_m }{\partial \mathsf K_k \partial \mathsf K_{k'}}  & = 0\\
    \frac{\partial \beta_m}{\partial \mathsf K_k } & = \mathsf X_k^{(\varphi(m))} \label{dbeta_dk} \\
    \frac{\partial^2 C }{\partial \mathsf K_k \partial \mathbf{g}^{(\omega)}} & = - \mathsf{X}_k^{(\omega)} \mathsf e^{(\omega)} \left( 2+ \left( \mathsf K^{\mathsf T} \mathsf X^{(\omega)} -\mathbf{g}^{(\omega)} \right) \vphantom{\frac{\mathbf g^{(\omega)}}{s^2}} \right.  \nonumber\\
    & \quad \cdot \left. \left( \frac{\mathbf{g}^{(\omega)}-\mathbf{g}_{\mathrm{lim}}'}{s^2} \right) \right) \\
    \frac{\partial^2 \beta_m }{\partial \mathsf K_k \partial \mathbf{g}^{(\omega)}} & = 0\\
    \frac{\partial \beta_m }{\partial \mathbf{g}^{(\omega)}} & = \frac{-2rM\exp{\left(-2r(\mathbf{g}_{\mathrm{lim}}+ \nu -\mathbf{g}^{(\omega)} )  \right)}}{ \left( 1+\exp{\left(-2r(\mathbf{g}_{\mathrm{lim}}+ \nu -\mathbf{g}^{(\omega)} )  \right)} \right)^2 } \label{dbeta_dg},
\end{align}
\end{subequations}
where $\varphi(m)\in \Omega$ maps the active constraint index $m$ to the corresponding index in the data set; $\mathsf e^{(\omega)}$ and $\mathbf{g}_{\mathrm{lim}}'$ are defined as follows:
\begin{subequations}
\label{e&g}
\begin{align}
    \mathsf e^{(\omega)} & = \exp{\left(-\frac{(\mathbf{g}^{(\omega)}-\mathbf{g}_{\mathrm{lim}}')^2}{2s^2}\right)}\\
    \mathbf{g}_{\mathrm{lim}}' & = \mathbf{g}_{\mathrm{lim}} + \frac{\nu}{2}.
\end{align}
\end{subequations}
Note that \eqref{dbeta_dk} and \eqref{dbeta_dg} are derived based on \eqref{coef_ctr3_cont} and the results associated with \eqref{coef_ctr2_cont} can be obtained in a very similar manner, thus not being covered here. 

\subsection{Sensitivity of Stability Index with Respect to System Parameters} \label{sec:3.4}
A number of stability indices are considered and discussed within the proposed stability-constrained system optimization framework in \cite{Part_I}, the sensitivity of these stability indices with respect to the uncertain parameters can be further derived analytically. An example that involves both matrix inverse and eigenvalue operations is given here to illustrate the concept. Consider a power system having $b\in\mathcal{B}$ buses with $g\in\mathcal{G}$, $c_l\in\mathcal{C}_l$ and $c_m\in\mathcal{C}_m$ being the set of conventional SGs, GFL and GFM IBRs. $\Psi(g)$ and $\Phi(c)$ map the units in $g\in \mathcal{G}$ and $c\in \mathcal{C}=\mathcal{C}_l\cup\mathcal{C}_m$ to the corresponding bus indices respectively. The system stability constraint can be expressed as follows \cite{gOSCR}:
\begin{align}
    \mathbf{g} (Y_{red}(\mathsf X, \mathsf p), \mathsf X) & = \lambda_{\mathrm{min}} \underbrace{\left[   \mathrm{diag}\left(\frac{V^2_{\Phi(c_l)}}{P_{c_l}}\right) Y_{red}(\mathsf X, \mathsf p) \right]}_{\mathsf Y_{red}'(\mathsf X, \mathsf p)} \nonumber \\
    & \ge \mathrm{gSCR}_{\mathrm{lim}} \label{Y_red'},
\end{align}
where $\mathrm{diag}\left({V^2_{\Phi(c_l)}}/{P_{c_l}}\right)$ is the diagonal matrix related to the GFL IBR terminal voltage ($V^2_{\Phi(c_l)}$) and output power ($P_{c_l}$) and $Y_{red}$ is the reduced node admittance matrix after eliminating passive buses and infinite buses. It has been revealed that the smallest eigenvalue of $Y_{red}'$ represents the connectivity of the network, and thus the grid voltage strength \cite{gOSCR}.

Since the derivative with respect to the uncertain parameters ($\partial \mathbf{g}/\partial \mathsf p $) are of concern, the decision variable $\mathsf X$ is omitted for the rest of the derivation, i.e., $\mathbf{g} = \mathbf{g} (Y_{red}(\mathsf p))$. $Y_{red}(\mathsf p)$ can be further expressed as: 
\begin{equation}
\label{Y_red}
    Y_{red}(\mathsf p) = Y_{\mathcal{C}_L\mathcal{C}_L}- Y_{\mathcal{C}_L \delta}Y_{\delta \delta}^{-1}(\mathsf p) Y_{\delta \mathcal{C}_L},
\end{equation}
where the terms on the right-hand side are the sub-matrices of the admittance matrix ($Y$) defined by:
\begin{equation}
    Y=\begin{bmatrix}
        \begin{array}{c|c}
        Y_{\mathcal{C}_L\mathcal{C}_L} & Y_{\mathcal{C}_L \delta} \\ \hline
        Y_{\delta \mathcal{C}_L} & Y_{\delta \delta} (\mathsf p)
        \end{array}
    \end{bmatrix}.
\end{equation}
with $\mathcal{C}_L\in\mathcal{B}$ being the set of GFL IBR nodes and $\delta = \mathcal{B}\setminus \mathcal{C}_L$ being the set of the rest nodes. The partial derivative $\partial \mathbf{g}/\partial \mathsf p$ can be derived by first applying the eigenvalue perturbation theorem:
\begin{align}
\label{eigen_perturb}
    \frac{\partial \mathbf{g}}{\partial \mathsf p_p} & = \frac{w^{\mathsf T}\frac{\partial Y_{red}'(\mathsf p)}{\partial \mathsf p_p}v}{w^{\mathsf T} v} 
\end{align}
where $w$ and $v$ are the left and right eigenvectors of the matrix $Y_{red}'(\mathsf p)$ corresponding to $\lambda_{\mathrm{min}}$. The expression of $\frac{\partial Y_{red}'(\mathsf p)}{\partial \mathsf p_p}$ can be further derived by combining \eqref{Y_red'} and \eqref{Y_red}:
\begin{align}
\label{derivative_Y_red'}
    \frac{\partial Y_{red}'(\mathsf p)}{\partial \mathsf p_p} & = - \mathrm{diag}\left(\frac{V^2_{\Phi(c_l)}}{P_{c_l}}\right) Y_{\mathcal{C}_L \delta} Y_{\delta \delta}^{-1} \frac{\partial Y_{\delta \delta} (\mathsf p)}{\partial \mathsf p_p} Y_{\delta \delta}^{-1}  Y_{\delta \mathcal{C}_L}.
\end{align}
To derive $\frac{\partial Y_{\delta \delta} (\mathsf p)}{\partial \mathsf p_p}$, the formula of the admittance matrix $Y$ is revisited:
\begin{align}
\label{Y}
    Y &= Y^0 +  Y^g,
\end{align}
where $Y^0$ is the admittance matrix of the transmission lines only; $Y^g$ denotes the additional $Y$ matrix increment due to the reactance of SGs and GFM IBRs. The element of the $i$-th row and $j$-th column in $Y^g$ can be expressed as below:
\begin{equation}
\label{Y2}
    Y_{ij}^g=
    \begin{cases}
    \frac{1}{X_{g}}x_{g}\;\;&\mathrm{if}\,i = j \land \exists\, g\in \mathcal{G}\cup \mathcal{C}_m,\, \mathrm{s.t.}\,i=\Psi(g)\\
    0\;\;& \mathrm{otherwise}.
    \end{cases}
\end{equation}
Considering the uncertainty of SG's and GFM IBR's reactance, i.e., $\mathsf p_p = X_g,\,g\in \mathcal{G}\cup \mathcal{C}_m$, the concerned derivative is derived as:
\begin{equation}
\label{derivative_Y_dd}
    \frac{\partial Y_{\delta \delta, ij}}{\partial X_g} = \begin{cases}
    -\frac{x_{g}}{X_{g}^2}\;\;&\mathrm{if}\,i = j \land \exists\, g\in \mathcal{G}\cup \mathcal{C}_m,\, \mathrm{s.t.}\,i=\Psi(g)\\
    0\;\;& \mathrm{otherwise}.
    \end{cases}
\end{equation}
Combining \eqref{eigen_perturb}, \eqref{derivative_Y_red'} and \eqref{derivative_Y_dd} leads to the closed form expression of ${\partial \mathbf{g}}/{\partial \mathsf p_p}$. It should be noted that although the uncertainty of the SG and GFM IBR reactance is considered in the derivation, other uncertain parameters, such as line reactance can be dealt with similarly. Moreover, the sensitivity of other stability indices, such as small-signal stability indices based on the eigenvalue of the closed loop system state matrix or post-fault current and voltage and different forms of system strength based on the elements in the impedance matrix can be developed in a similar yet simpler manner. 

With the derivation in Section~\ref{sec:3.3} and Section~\ref{sec:3.4}, the first order derivatives $\nabla f$ can be computed according to \eqref{chain_1}. As for the second derivatives $\partial^2 f_k/\partial \mathsf p_p^2$, although the analytical derivation may become extremely cumbersome due to the complex expressions in $\nabla f$, numerical approaches with good approximation can be applied \cite{strikwerda2004finite}. As a result, the statistical moments of the stability constraint coefficients can be calculated based on \eqref{E_K} and \eqref{Cov_K}.

\section{Distributionally Robust Stability-Constrained Optimization} \label{sec:4}
Having obtained the uncertainty information of the stability constraint coefficients, a distributionally robust stability-constrained UC problem can be formulated, where the overall system operation cost is minimized subjected to a number of constraints, such as power flow and power balance constraints, thermal unit constraints, and the distributionally robust system stability constraints.

\subsection{Objective Function} \label{sec:4.1}
The objective of the UC problem is to minimize the expected cost over all nodes in the given scenario tree:
\begin{equation}
    \label{eq:SUC}
    \min \sum_{n\in \mathcal{N}} \pi (n) \left( \sum_{g\in \mathcal{G}}  C_g(n) + \Delta t(n) c^s P^s(n) \right)
\end{equation}
where $\pi(n)$ is the probability of scenario $n\in \mathcal{N}$ and $C_g(n)$ is the operation cost of unit $g\in \mathcal{G}$ in scenario n, including startup, no-load and marginal cost; $\Delta t(n)c^sP^s(n)$ represents the cost of the load shedding in scenario n with the three terms being the time step of scenario n, load shedding cost and shed load. The scenario tree is built based on user-defined quantiles of the forecast error distribution to capture the uncertainty associated with demand and wind generation. \textcolor{black}{Section~II in \cite{7115982} can be referred to for more details.}

\subsection{Distributionally Robust Stability Constraints}
Assume the first- and second-order moments of the SG and GFM impedance are known, denoted by $\mu_{\mathsf p}$ and $\Sigma_{\mathsf p}$ respectively, whereas the exact probability distribution is unknown. Based on the discussion in Section~\ref{sec:3}, the first- and second-order moments of the coefficients of the stability constraint can be developed and are denoted by $\mu_{\mathsf K}$ and $\Sigma_{\mathsf K}$ respectively. As a result, the ambiguity set of the coefficients $\mathsf K$ can be modeled as:
\begin{align}
\label{ambiguity_set}
    \mathcal{P} =\Big\{\mathbf{D} \in \xi(\mathsf K):\; & \mathbb{E}^\mathbf{D}(\mathsf K) = \mu_{\mathsf K}, \mathrm{Var}^\mathbf{D}(\mathsf K) =\Sigma_{\mathsf K} \Big\}
\end{align}
where $\xi(\cdot)$ is the probability density function; $\mu_{\mathsf K}$ and $\Sigma_{\mathsf K}$ denote the mean and covariance matrix of the uncertain parameters given the distribution $\mathbf{D}$. 

With the above ambiguity set, the distributionally robust stability chance constraint can be formulated as follows: 
\begin{equation}
\label{Robust_Stability}
    \min_{\mathbf{D}\in\mathcal{P}} \mathrm{Pr} \left\{ -\mathsf K^{\mathsf T} \mathsf X \le -\mathbf{g}_{\mathrm{lim}}    \right\} \ge \eta,
\end{equation}
which maintains the stability constraint over a certain confidence level $\eta\in (0,1)$, for all the possible distribution in the ambiguity set $\mathcal{P}$. According to \textit{Theorem 3.1} in \cite{calafiore_ghaoui_2022}, the above distributionally robust chance constraint can be equivalently converted to:
\begin{subequations}
\begin{align}
\label{Robust_Stability_SOC}
    k_{\eta} \sqrt{\mathsf X^{\mathsf T} \Sigma_{\mathsf K} \mathsf X}  & \le \mu_{\mathsf K}^{\mathsf T} \mathsf X -\mathbf{g}_{\mathrm{lim}} \\
    \hspace{1.96cm} k_\eta & = \sqrt{\frac{\eta}{1-\eta}}.
\end{align}
\end{subequations}
\textcolor{black}{It should be noted that although the results in \eqref{Robust_Stability_SOC} tend to be conservative, this conservativeness does not come from the Chebychev inequality itself, since \eqref{Robust_Stability} to \eqref{Robust_Stability_SOC} is an equivalent transition with no approximations as shown in \cite{calafiore_ghaoui_2022}. The conservativeness is due to the moment-based ambiguity set in \eqref{Robust_Stability}, which requires the probability that the stability constraint holds to be greater than $\eta$ for all the distributions with the given first- and second-order moments. In reality, if more information regarding the distribution is known, the conservativeness can be reduced significantly. For instance, if the uncertain parameter is centrally symmetric around its mean, the coefficient $k_{\eta}$ in (39b) is reduced from $\sqrt{\frac{\eta}{1-\eta}}$ to $\sqrt{\frac{1}{2(1-\eta)}}$, $\forall \eta \in [0.5,1)$ \cite{calafiore_ghaoui_2022}.}

Due to the fact that the covariance matrix $\Sigma_{\mathsf K}$ is symmetric and positive semi-definite, it can be expanded by the spectral factorization:
\begin{equation}
\label{spectral}
    \Sigma_{\mathsf K} = \sum_i \tau_i q_i q_i^{\mathsf T},
\end{equation} 
with $\tau_i$ and $q_i$ being the $i$-th eigenvalue and orthogonal eigenvector respectively. Substituting \eqref{spectral} into \eqref{Robust_Stability_SOC} gives the following stability constraint:
\begin{equation}
\label{SSS_SOC}
    \left\Vert \begin{bmatrix}
        \sqrt{\tau_1}q_1^{\mathsf T}\mathsf X \\
        \dots\\
        \sqrt{\tau_n}q_n^{\mathsf T}\mathsf X
    \end{bmatrix} \right\Vert_2 \le \frac{1}{k_{\eta}}\left(\mu_{\mathsf K}^{\mathsf T} \mathsf X -\mathbf{g}_{\mathrm{lim}}\right).
\end{equation}
With $\tau_i\ge 0$, \eqref{SSS_SOC} is a well-defined constraint in a standard SOC form.

All other conventional UC constraints such as those of power balance, thermal units and transmission system are not listed in the paper. \textcolor{black}{The readers can refer to (5)-(7) in \cite{Part_II} for more details.}

\section{Discussion on An Alternative Formulation} \label{sec:5}
In previous sections, the parameter uncertainties related to the system dynamics, are transferred to the coefficients of the stability constraints in the system-level optimization. In this way, the uncertainty associated with the system dynamics is essentially managed by a distributional robust chance-constrained formulation during the system scheduling process. Alternatively, this uncertainty can also be explicitly considered when generating the stability constraint coefficients. 

For this purpose, the formulation based on the concept of distributionally robust learning is discussed here. With this formulation, the training process determines the optimal coefficients of the stability constraints that best describe the relationship between the stability index and the decision variables, while considering the uncertainty in the data set. Given the notation in \eqref{DM_compact}, a general formulation can be expressed as:
\begin{subequations}
\label{DR_II}
    \begin{align}    
    \min_{\mathsf K} \quad &\max_{\mathbf{D}_{\mathsf p} \in \mathcal{P}_{\mathsf p}} \mathbb E^{\mathbf{D}_{\mathsf p}} \left[ C(\mathsf K, \mathbf{g}^{\Omega}(p))\right] \\
    \mathrm{s.t.} \quad & \min_{\mathbf{D}_{\mathsf p} \in \mathcal{P}_{\mathsf p}} \mathrm{Pr}  \left\{\beta (\mathsf K, \mathbf{g}^{\Omega}(p)) \ge 0 \right\} \ge \eta,
    \end{align}    
\end{subequations}
where $\mathbf{D}_{\mathsf p}$ and $\mathcal{P}_{\mathsf p}$ are the the distribution of the uncertain vector $\mathsf p$ and the corresponding ambiguity set. After obtaining the optimal stability constraint coefficients $\mathsf K$, the system scheduling can be modeled as a standard optimization problem with a simple stability constraint \eqref{linear_glim}. It is theoretically possible to reformulate problem \eqref{DR_II} into a tractable convex model for simple linear regression or more general cases, where the objective function and the constraints depend affinely on the random vector $\mathsf p$ \cite{zymler2013distributionally,calafiore_ghaoui_2022}. However, considering the highly nonlinear relationship in $C(\mathbf{g}^{\Omega})$, $\beta(\mathbf{g}^{\Omega})$ and $\mathbf{g}^{\Omega}(\mathsf p)$, as defined in the constrained and weighted regression problem \eqref{DM3} or \eqref{DM_compact}, \eqref{DR_II} becomes computationally intractable. Moreover, another drawback with this formulation is the repeated data generation and retraining process, every time the uncertainty information of $\mathsf p$ and the confidence level requirement $\eta$ change. Nevertheless, this work focuses on the formulation developed in previous sections and an efficient reformulation of \eqref{DR_II} and its comparison against this work will be conducted in future work.

\section{Case Studies} \label{sec:6}
\begin{figure}[!t]
    \centering
    \vspace{-0.35cm}
	\scalebox{0.42}{\includegraphics[trim=0 0 0 0,clip]{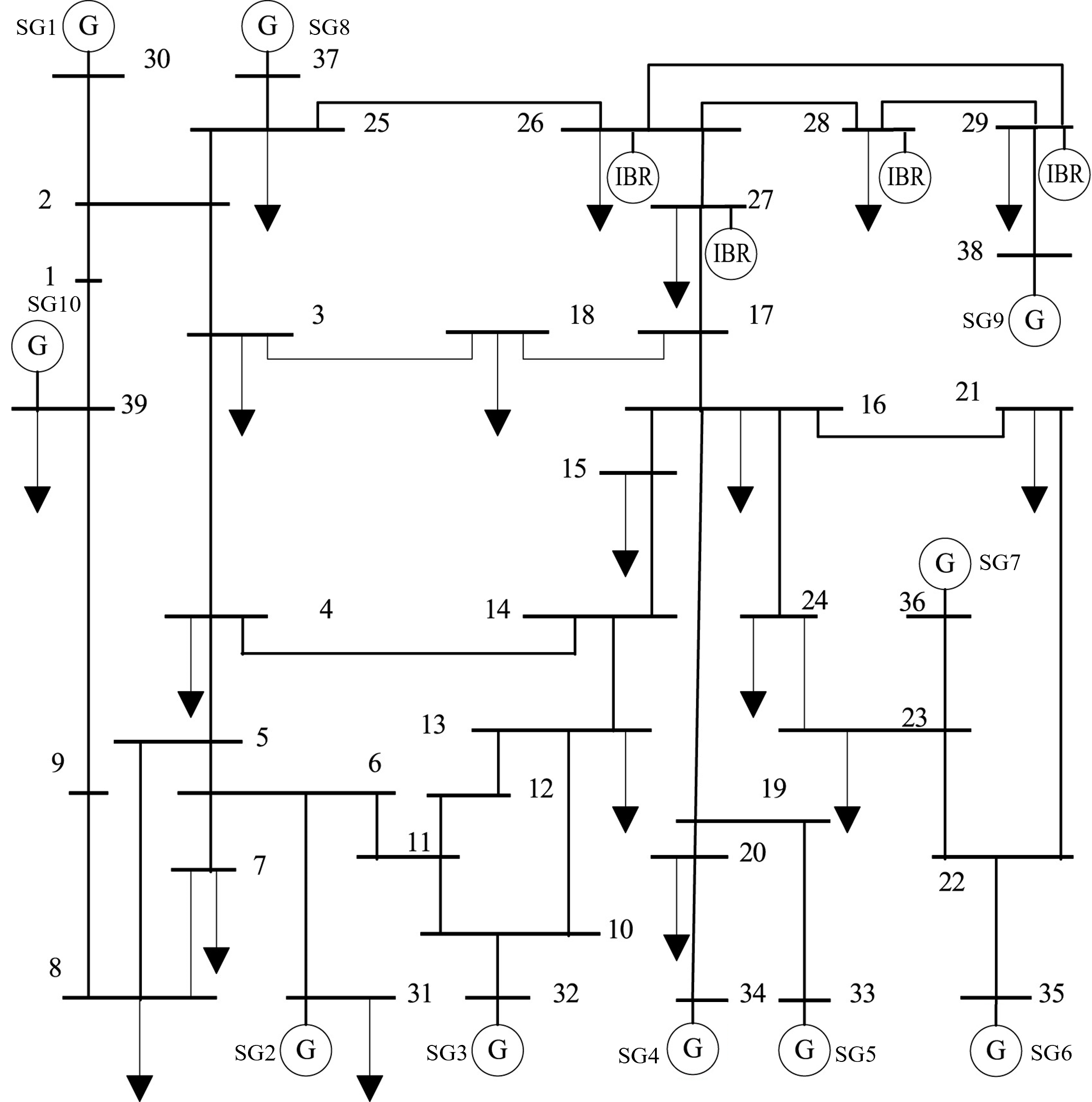}}
    \caption{\label{fig:39-bus}Modified IEEE-39 bus system.}
    \vspace{-0.35cm}
\end{figure}

The IEEE-39 bus system shown in Fig.~\ref{fig:39-bus} is utilized to demonstrate the effectiveness of the proposed model. To increase the renewable penetration, IBRs are added at Bus 26, 27, 28 and 29 with the second one being a GFM unit.
The parameters of transmission lines and loads are available in \cite{39_bus}. The load and renewable generation profile in \cite{9968474} is adapted for the simulation during the considered time horizon. The characteristics of the thermal generators are given in Table~\ref{tab:SG_para}
% while considering the data in \cite{9968474}, 
with their location being Bus $\{30,37\}$, $\{31,32,33,34,35,36,38\}$ and $\{39\}$ respectively. Other system parameters are set as follows: load demand $P^D\in [5.16, 6.24]\,\mathrm{GW}$, load damping $D = 0.5\% P^D / 1\,\mathrm{Hz}$, base power $S_B = 100\mathrm{MVA}$. \textcolor{black}{The MISOCP-base UC problem is solved by Gurobi (10.0.0) on a PC with Intel(R) Core(TM) i7-7820X CPU @ 3.60GHz and RAM of 64 GB.}

\textcolor{black}{\textbf{Discussion on the training data generation:} For the considered index given by \eqref{Y_red'}, the decision variables (before the augmentation) are the status of SGs ($x_g$) and IBRs ($P_{c_l}$). For the SGs, all the possible generator combinations are considered, whereas for the continuous decisions $P_{c_l}\in [0,1] \,\mathrm{p.u.}$, there are infinite possible conditions in theory. To obtain the data set with a finite size, the interval is evenly divided into $n$ regions, each of which is represented by its mean value. In practice, different approaches can be considered to reduce the number of training samples, such as the enforcement of the power balance constraint for a certain time horizon and bus aggregation. Additionally, there are also efficient data generation approaches in the literature such as \cite{thams2019efficient} for data-driven stability assessment that can be potentially utilized to increase the efficiency of data generation. Since the instability is mainly caused by the lack of system strength, the unstable samples can be generated by reducing the number of online SGs. For a large system with significant variation in the demand and renewable generation, the regression model \eqref{DM3} may need to be solved on a weekly/monthly basis. Nevertheless, since optimization \eqref{DM3} is a simple linearly constrained quadratic problem, solving it requires only a few seconds.}

\textcolor{black}{\textbf{Discussion on decision augmentation:} The augmented decision variables, $\mathsf X$ in \eqref{g_SOC} are designed to increase the representability of the regression model. Specifically, the pairwise products of any two decisions are used to account for the interaction of these two units. Although this augmentation may lead to a significant amount of decisions in large systems, it is not necessary to include all the combinations in practice. On one hand, it is understandable that the two units that are electrically far away from each other have little interaction in terms of stability performance. Hence, the products between those units are unnecessary. On the other hand, after solving the regression model, there exist augmented variables whose coefficients are much (at least one order of magnitude) less than those of the others, meaning that the impact of these terms on the stability index is insignificant, thus being neglected. Note that the training process is repeated after the elimination of the insignificant terms. As for the case study, the vector $\mathsf X$ includes 1 ($\mathsf X_1\in\mathbb R^1$), the online status of SGs and GFMs ($\mathsf X_2\in\mathbb R^{11}$), the wind power in the system ($\mathsf X_3\in\mathbb R^1$) and the product of every $\mathsf X_2$ and $\mathsf X_3$ ($\mathsf X_4\in\mathbb R^{11}$), leading to a final decision of $\mathsf X\in \mathbb R^{24}$. }

\begin{table}[!t]
\renewcommand{\arraystretch}{1.2}
\caption{{Parameters of Thermal Units}}
\label{tab:SG_para}
\noindent
\centering
    \begin{minipage}{\linewidth} %Use the minipage environment to footnote tables
    \renewcommand\footnoterule{\vspace*{-5pt}} %to remove the horizontal rule above the table footnote
    \begin{center}
        \begin{tabular}{ c || c | c | c }
            \toprule
            Type & \textbf{Type I} & \textbf{Type II} & \textbf{Type III} \\ 
            \cline{1-4}
            % Bus number & 31 & 32 & 33 & 34 & 35 & 39 \\ 
            % \cline{1-7}
            % Rated Power [MW]& 678 & 650 & 632 & 508 & 650 & 1000\\
            % \cline{1-7} 
            % Min Stable Gen [MW] & 330 & 225 & 306 & 204 & 225 & 500  \\
            % \cline{1-7} 
            No-load Cost [k\pounds/h]& 4.5 & 3 & 0\\
            \cline{1-4} 
            Marginal Cost [\pounds/MWh]& 47 & 200 & 10\\
            \cline{1-4} 
            Startup Cost [k\pounds]& 10 & 0 & N/A \\
            \cline{1-4} 
            Startup Time [h]& 4 & 0 & N/A\\
            \cline{1-4}
            Min Up Time [h] & 4 & 0 & N/A\\
            \cline{1-4} 
            Min Down Time [h] & 1 & 0 & N/A \\
            \cline{1-4}
            Inertia Constant [s]& 6 & 6 & 6 \\
           \bottomrule
        \end{tabular}
    \end{center}
    \end{minipage}
    \vspace{-0.35cm}
\end{table} 

\subsection{Validation of Uncertainty Quantification} \label{sec:6.1}
\begin{figure}[!t]
    \centering
    \vspace{-0.35cm}
	\scalebox{1.15}{\includegraphics[]{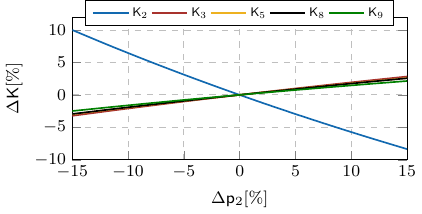}}
    \caption{\label{fig:Linear}{\textcolor{black}{Relationship between stability constraint coefficients and uncertain parameters}.}}
    \vspace{-0.35cm}
\end{figure}

To validate the effectiveness of the uncertainty quantification method proposed in Section~\ref{sec:3}, the mean and variance of the stability constraint coefficients obtained based on the following two methods are compared: i) the proposed analytical (Ana.) method and ii) Monte Carlo (MC) simulation with the following steps. \textcolor{black}{First, generate a sufficient amount of samples (15k in this case to ensure the convergence), such that the mean and variance of the uncertain parameters are equal to the given values. However, it is almost impossible to generate samples from all possible distributions. Since we are only interested in the moments of the uncertain parameters, a normal distribution is assumed without loss of generality. The higher moments have little impact on the proposed method as long as the higher derivatives are negligible according to the derivation in Section~\ref{sec:3.2}, which is the case for our application in this work as assessed here. The relationship between the stability constraint coefficients and the uncertain parameter is depicted in Fig.~\ref{fig:Linear}. An approximately linear relationship between the uncertain parameter and the stability constraint coefficients can be observed. This linear relationship can be partially explained by the regression model with only linear constraints. Note that the normalized variation is used such that the relationship can be compared within one figure. Although only the relationship between 5 coefficients with respect to one uncertain parameter is depicted to avoid overlapping, all the others present a very similar trend. Second, for each sample, train the regression model \eqref{DM_compact} that gives a sample of the stability constraint coefficients. Finally, calculate the mean and variance of the stability constraint coefficients.}

From the results listed in Table~\ref{tab:error}, it can be observed that both the mean ($\mu$) and variance ($\sigma^2$) derived based on the analytical approach are close to those obtained from MC simulation. \textcolor{black}{The mean absolute percentage error (MAPE) of the mean and variance are 5.04 \% and 7.82 \% respectively being in comparable range with that in uncertainty quantification research \cite{ye2021uncertainty,tang2015dimension,qiu2020nonintrusive}, where the uncertainty of different quantities such as load and renewables are quantified with errors being in the range of a few to around 10 percent are reported, demonstrating a good approximation of the proposed method.}
\begin{table}[!t]
\renewcommand{\arraystretch}{1.2}
\caption{{Mean and Variance of Stability Constraint Coefficient with Different Methods}}
\label{tab:error}
\noindent
\centering
    \begin{minipage}{\linewidth} %Use the minipage environment to footnote tables
    \renewcommand\footnoterule{\vspace*{-5pt}} %to remove the horizontal rule above the table footnote
    \begin{center}
        \begin{tabular}{c||ccc|ccc}
            \toprule
            \multirow{2}{*}{} & \multicolumn{3}{c|}{\pmb{$\mu$}} & \multicolumn{3}{c}{\pmb{$\sigma^2$} [$\mathrm{\times 10^{-3}}$]}        \\ 
            \cline{2-7} 
            & \multicolumn{1}{c|}{\textbf{Ana.}} & \multicolumn{1}{c|}{\textbf{MC}} & \multicolumn{1}{c|}{\pmb{$e_{\mu}$} [$\mathrm{\%}$]} & \multicolumn{1}{c|}{\textbf{Ana.}} & \multicolumn{1}{c|}{\textbf{MC}} & \multicolumn{1}{c}{\pmb{$e_{\sigma^2}$} [$\mathrm{\%}$]} \\ 
            \hline
            $\mathsf K_1$                & \multicolumn{1}{c|}{2.392}   & \multicolumn{1}{c|}{2.496}    &  -4.19  & \multicolumn{1}{c|}{0.438} &  \multicolumn{1}{c|}{0.405} & 8.05 \\ 
            \hline
            $\mathsf K_2$                & \multicolumn{1}{c|}{1.458}   &    \multicolumn{1}{c|}{1.474}   &  -1.09   & \multicolumn{1}{c|}{3.470} &  \multicolumn{1}{c|}{3.644} & -4.78       \\ 
            \hline
            $\mathsf K_3$                & \multicolumn{1}{c|}{0.407}   &   \multicolumn{1}{c|}{0.389}    &  4.65  & \multicolumn{1}{c|}{0.399} &  \multicolumn{1}{c|}{0.392} & 1.97       \\ 
            \hline
            $\mathsf K_4$                & \multicolumn{1}{c|}{0.478}   &    \multicolumn{1}{c|}{0.471}   &  1.56  & \multicolumn{1}{c|}{0.249} &  \multicolumn{1}{c|}{0.258} & -3.25       \\ 
            \hline
            $\mathsf K_{5}$                & \multicolumn{1}{c|}{-0.078}   &    \multicolumn{1}{c|}{-0.088}    &  -11.96 & \multicolumn{1}{c|}{0.689} &  \multicolumn{1}{c|}{0.653} & 5.51       \\ 
            \hline$\mathsf ...$    & \multicolumn{1}{c|}{...}   &   \multicolumn{1}{c|}{...}     & ...  & \multicolumn{1}{c|}{...}   &   \multicolumn{1}{c|}{...}    &...    \\ 
            \hline
            $\mathrm{\textcolor{black}{MAPE}}$              & \multicolumn{1}{c|}{N/A}   &   \multicolumn{1}{c|}{N/A}    & 5.04   &    \multicolumn{1}{c|}{N/A}  & \multicolumn{1}{c|}{N/A}   &   7.82   \\
            \bottomrule
        \end{tabular}
    \end{center}
    \end{minipage}
    \vspace{-0.35cm}
\end{table} 

\begin{figure}[!t]
    \centering
    % \vspace{-0.35cm}
	\scalebox{0.5}{\includegraphics[]{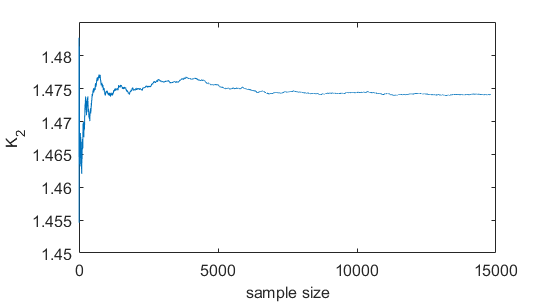}}
    \caption{\label{fig:MC}{\textcolor{black}{Sample averages at different sample sizes}.}}
    \vspace{-0.35cm}
\end{figure}

\textcolor{black}{A sensitivity analysis of the approximation error with respect to the sample size and the variation ranges is also carried out. The sample average is calculated at different sample sizes with the results shown in Fig~\ref{fig:MC}. Good convergence can be observed after the sample size exceeds 10k or sometimes even 5k, which validates the sufficiency of the sample size in the case studies. Again, only the average of one parameter is selected due to the space limitation and those of the others present a similar trend. Simulations are also carried out with different Coefficients of Variation (CV), which is defined as the ratio of the standard deviation to the mean. The results are listed in Table~\ref{tab:range}. It can be observed that as the variation range increases the mean absolute percentage errors against the MC results of both mean and variance grows, which is understandable due to the approximation based on the Taylor expansion. However, even in the case where the CV equals 20\%, the approximation error is still within an acceptable range. On the other hand, considering the actual uncertainty range in power systems \cite{9020274,karayaka2003synchronous}, the practical parameter uncertainty range of generators is typically within 10\%, which also illustrates the effectiveness of the proposed method in practical implementation.} 

\begin{table}[!b]
% \vspace{-0.35cm}
\renewcommand{\arraystretch}{1.2}
\caption{\textcolor{black}{Approximation under different Coefficients of Variation (CV)}}
\label{tab:range}
\noindent
\centering
    \begin{minipage}{\linewidth} %Use the minipage environment to footnote tables
    \renewcommand\footnoterule{\vspace*{-5pt}} %to remove the horizontal rule above the table footnote
    \begin{center}
        \begin{tabular}{ c || c | c | c | c }
            \toprule
            CV [\%] & \textbf{5} & \textbf{10} & \textbf{15} & \textbf{20} \\ 
            \cline{1-5}
            $\mathrm{MAPE}_{\mu}$ & 5.04 & 6.01 & 7.25 & 8.67\\
            \cline{1-5} 
            $\mathrm{MAPE}_{\sigma^2}$ & 7.82 & 8.02 & 8.26 & 8.73\\
           \bottomrule
        \end{tabular}
    \end{center}
    \end{minipage}
\end{table} 

\textcolor{black}{Note that the comparison against some other existing approaches is not straightforward or well-defined due to the following reasons. i) The concerned problem of the parameter uncertainty associated with the system dynamic model within the framework of stability-constrained optimization has not been discussed or dealt with in the literature. ii)~The comparison with other approaches such as robust and chance-constrained optimization may be unnecessary, since the authors do not claim the formulation based on the DRO in the presented method is the best choice under all circumstances. Actually, applying which type of uncertainty management approach depends on the system operators' knowledge of the uncertain parameter as discussed in Section~\ref{sec:3.1}.}

\subsection{Impact of Stability Constraints on System Operation} \label{sec:6.2}
The effectiveness of the proposed stability constraints and their influences on the system operation cost as well as operating conditions are studied in this section. The following three different cases are considered: 
\begin{itemize}
    \item Case I: conventional system scheduling without stability constraints
    \item Case II: system scheduling with stability constraints and the system parameter uncertainty is not considered
    \item Case III: system scheduling with distributionally robust stability constraints to account for system uncertainties.
\end{itemize}

\subsubsection{\textbf{System operation cost and stability constraint violation}} 
\begin{figure}[!t]
    \centering
    \vspace{-0.35cm}
	\scalebox{1.15}{\includegraphics[]{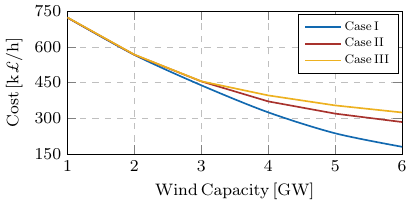}}
    \caption{\label{fig:Cost_wind}{System operation costs with varying wind capacity.}}
    \vspace{-0.35cm}
\end{figure}

The averaged system operation costs among 24-hour operation in these three cases with various wind penetration levels are depicted in Fig.~\ref{fig:Cost_wind}. It is understandable that the system operation costs in all the cases decrease, as more wind generation is installed in the system. These costs stay close to each other at lower wind capacity (below $3\,\mathrm{GW}$), since the stability issue is not obvious. However, as the installed wind capacity in the system further increases, the system operation cost in Case I (blue curve) presents the lowest value among all the cases, because the stability constraint is not considered in the scheduling process. Once the stability constraint is incorporated into the system scheduling, the original optimal solution cannot be attained, thus inducing more operational cost to maintain the stability constraint, as indicated by the red curve. This cost increment becomes more significant at high wind capacity due to the server stability issue. In addition, if the uncertain associated with the system parameters is considered (Case III), the system operation cost further increases by around $40\, \mathrm{k\pounds/h}$ at $6\,\mathrm{GW}$ wind capacity, since a larger stability margin has to be reserved to ensure the stability under uncertainty.

\begin{figure}[!b]
    \centering
    \vspace{-0.35cm}
	\scalebox{1.15}{\includegraphics[]{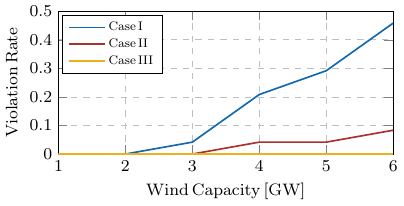}}
    \caption{\label{fig:VR}{Stability constraint violation rates in different cases with varying wind capacity.}}
    \vspace{-0.35cm}
\end{figure}
The stability constraint violation rates under different circumstances are also shown in Fig.~\ref{fig:VR}. In Case I, as indicated by the blue curve, the stability constraint violation starts to appear at $3\,\mathrm{GW}$ wind capacity for about 5\% of the total operation time. This violation rate increases dramatically to about 45\%, with the rising of the wind capacity in the system, which demonstrates the necessity of including the stability constraint in the system optimization. As for Case II, although the stability constraint is considered, a small percentage (less than 10\%) of the operating conditions do not satisfy the stability constraint due to the neglect of the parameter uncertainty related to system dynamics. This impact becomes more remarkable at higher wind capacity, illustrating the importance of modeling and management of the uncertainty in system stability constraints. The stability constraint violation due to the uncertainty in Case II is eliminated in Case III with the incorporation of the distributionally robust stability chance constraint, therefore demonstrating the effectiveness of the proposed method. \textcolor{black}{Note that it is unlikely to quantify the real-time adjustment costs due to the stability violation as in the conventional UC and dispatch framework that deals with the uncertainty of the renewables and demand, as the uncertainty associated with the system dynamic parameters does not reveal itself as time evolves.}

\begin{figure}[!t]
    \centering
    \vspace{-0.35cm}
	\scalebox{1.15}{\includegraphics[]{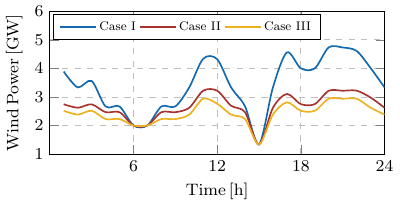}}
    \caption{\label{fig:WindOnline}{Dispatched wind power during 24-hour operation.}}
    \vspace{-0.35cm}
\end{figure}
\begin{figure}[!b]
    \centering
    \vspace{-0.35cm}
	\scalebox{1.15}{\includegraphics[]{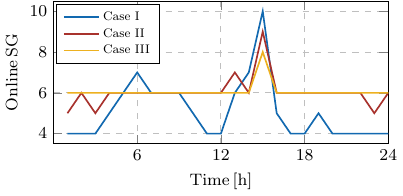}}
    \caption{\label{fig:SGOnline}{Number on online SGs during 24-hour system operation.}}
    \vspace{-0.35cm}
\end{figure}

\subsubsection{\textbf{Dispatched wind power and SGs}} In order to investigate the reasons behind the changes in the system operation costs and stability constraint violation rates in different cases, the amount of dispatched wind power and the number of online SGs during 24-hour operation are plotted in Fig.~\ref{fig:WindOnline} and Fig.~\ref{fig:SGOnline} respectively. The wind capacity in both figures are set to be $6\,\mathrm{GW}$. It can be observed in Fig.~\ref{fig:WindOnline}, that when no stability constraint is considered during the system scheduling process (Case I), the dispatched wind power, indicated by the blue curve is the highest during most of the hours among all the cases, which also justifies the lowest operation cost in this case as illustrated in Fig.~\ref{fig:Cost_wind}. Furthermore, since the stability constraint considered in \eqref{Y_red'} is closely related to the output power of the grid-following IBRs in the system as explained in \cite{gOSCR}, a significant amount of wind power is curtailed in Case II to ensure the system stability, especially during the high wind power hours. This effect becomes more obvious in Case III (yellow curve), when the uncertainty of the system dynamics is considered through the distributionally robust formulation. In this scenario, additional wind power (around $0.3\,\mathrm{GW}$) compared to Case II is further curtailed, which leads to a zero stability constraint violation rate under uncertainty as shown in Fig.~\ref{fig:VR}.

On the other hand, the impact on the number of online SGs is also investigated with the results plotted in Fig.~\ref{fig:SGOnline}. In general, the committed SGs in Case I are the smallest for most of the time since more wind power can be utilized without the stability constraint. Similarly, to provide enough system strength and ensure robustness against uncertainty, the number of SGs in Case III is larger than or equal to that in Case II. However, an opposite trend is observed during $6\,\mathrm{h}$ and $13-15\,\mathrm{h}$ when there is a significant decline in wind power. This is because in Case I, there is not enough reserve to balance the demand, from the SGs committed before the wind power decreases. As a result, more SGs have to be dispatched during these hours due to the SG ramp rate constraints. Therefore, the blue curve exceeds the red and yellow curves. Nevertheless, more SGs are needed on average to maintain system stability in the cases where the uncertainty and stability constraints are modeled and implemented in system scheduling.

\subsubsection{Comparison to operation with fixed stability margin}
\begin{figure}[!t]
    \centering
    \vspace{-0.35cm}
	\scalebox{1.1}{\includegraphics[]{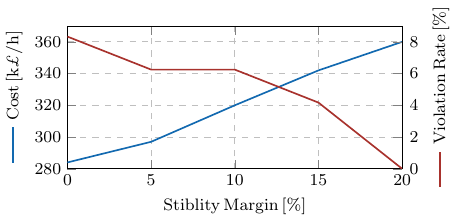}}
    \caption{\label{fig:SM}{System operation cost and constraint violation rate with different stability margin.}}
    \vspace{-0.35cm}
\end{figure}
\textcolor{black}{In order to illustrate the benefit of the proposed method, the operating strategy with a fixed stability margin to manage the uncertainty in the system is assessed, where the wind capacity in this case is set to
$6\,\mathrm{GW}$. As shown in Fig.~\ref{fig:SM}, a high stability margin leads to increased system operation cost and decreased stability constraint violation. A zero stability constraint violation requires a stability margin of $20\%$, which corresponds to the system operation cost of $360.84\,\mathrm{k\pounds/h}$, whereas with the proposed method, this number reduces to $324.65\,\mathrm{k\pounds/h}$. This indicates that the stability uncertainty is managed with an overconservative stability margin, since the dependence of the actual stability margin on the system operating conditions as expressed by the left-hand side in \eqref{SSS_SOC} cannot be considered in this strategy.}

\subsection{Impact of Uncertainty Level} \label{sec:6.3}
\begin{figure}[!t]
    \centering
    \vspace{-0.35cm}
	\scalebox{1.1}{\includegraphics[]{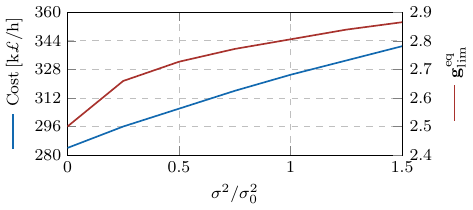}}
    \caption{\label{fig:Sigma}{System operation costs and the equivalent stability index limits under various uncertainty levels.}}
    \vspace{-0.35cm}
\end{figure}

The parameter uncertainties associated with the system dynamics are modeled and managed in this work through a distributionally robust formulation during the system scheduling process. The uncertainty level thus influences the optimal solution, which is investigated in this section. The results are depicted in Fig.~\ref{fig:Sigma}, where the averaged equivalent stability index limits ($\mathbf g_{\mathrm{lim}}^{\mathrm{eq}}$) is derived from \eqref{SSS_SOC} as follow:
\begin{equation}
    \mathbf g_{\mathrm{lim}}^{\mathrm{eq}} = \frac{1}{|\mathcal{T}|}\sum_{t\in\mathcal{T}}\left(\mathbf{g}_{\mathrm{lim}} + k_{\eta} \left\Vert \begin{bmatrix}
        \sqrt{\tau_1}q_1^{\mathsf T}\mathsf X(t) \\
        \dots\\
        \sqrt{\tau_n}q_n^{\mathsf T}\mathsf X(t)
    \end{bmatrix} \right\Vert_2 \right),
\end{equation}
representing the stability index limit under the uncertainty. It can be observed that more system operation cost is induced in an approximate linear fashion, as indicated by the blue curve, with the increasing of the uncertainty level, which is defined as the variance ratio to the original variance ($\sigma_0^2$). This can be explained by the robust formulation to ensure the stability constraint under different uncertainty levels. Note that $\sigma^2/\sigma_0^2 = 0$ corresponds to Case II in Section~\ref{sec:6.2} and $\sigma^2/\sigma_0^2 = 1$ Case III.

Similarly, a higher uncertainty level leads to a larger equivalent stability index limit, which represents a larger stability margin. To maintain this increased stability margin, more wind power is curtailed during high wind power hours, thus resulting in higher operational cost as discussed previously.

\color{black}
\subsection{Scalability}
The scalability of the proposed method is demonstrated through IEEE 118-bus system with the demand $[2000,5600]\,\mathrm{MW}$ and $S_B = 1000\,\mathrm{MVA}$. Wind generation is added at Bus 10, 26, 49, 61, 65 and 80 with a total capacity of $6\,\mathrm{GW}$. The generator parameters remain unchanged. The network data of the systems can be obtained in \cite{Data_system}.  

\begin{table}[!b]
\renewcommand{\arraystretch}{1.2}\caption{\textcolor{black}{Computational time of different cases and systems}}
\label{tab:time}
\noindent
\centering
    \begin{minipage}{\linewidth} %Use the minipage environment to footnote tables
    \renewcommand\footnoterule{\vspace*{-5pt}} %to remove the horizontal rule above the table footnote
    \begin{center}
        \begin{tabular}{ c || c | c | c | c | c }
            \toprule
              & \multicolumn{3}{c|}{$\mathbf{Time\,[s/step]}$} & \multicolumn{2}{c}{$\mathbf{MAPE\,[\%]}$} \\
             \cline{2-6}
             & Case I & Case II & Case III & \textbf{$\mu$} & \textbf{$\sigma^2$}\\ 
            \cline{1-6}
            \textbf{39-bus DUC} & 0.843 & 1.850 & 2.929 & \multirow{ 2}{*}{5.04} &\multirow{ 2}{*}{7.82}\\
            \cline{1-4}
            \textbf{39-bus SUC} & 1.234 & 12.661 & 17.079 & &\\
            \cline{1-6}
            \textbf{118-bus DUC} & 1.632 & 2.950 & 6.264 & 6.28 &8.33\\
           \bottomrule
        \end{tabular}
    \end{center}
    \end{minipage}
    \vspace{-0.3cm}
\end{table} 

The computational time of the three cases in Section~\ref{sec:6.2} in different test systems is reported in Table~\ref{tab:time}, where ``DUC'' and ``SUC'' represent deterministic and stochastic respectively. Note that it is the uncertainty of the renewable generation that is different in the DUC and SUC cases, as indicated in Section~IV-A. The computational time in Case II is increased due to the inclusion of the stability constraints. With further incorporation of the distributional robust formulation to manage the uncertainty, the computational time is not increased significantly compared with the increment due to the SUC formulation, thus highlighting the efficiency of the proposed method. It can be observed that although the computational time in the 118-bus system compared with the IEEE 39-bus system is increased, it is still within the acceptable range. Additionally, the approximation error is not increased significantly. As for the other results, due to their similarity to those in IEEE 39-bus system and the space limitation, they are not presented here.
\color{black}

\section{Conclusion and Future Work} \label{sec:7}
The uncertainty associated with the system dynamics is investigated in this work, within the framework of stability-constrained optimization in high IBR-penetrated systems. In order to ensure system stability under uncertainty, the uncertainty levels of the stability constraint coefficients are analytically quantified based on the uncertainty information related to the system impedance. A distributionally robust stability chance constraint is further formulated and converted into an SOC form, leading to an overall MISOCP-based system scheduling model. A good approximation of the uncertainty quantification method through moments propagation is observed and the impacts of the stability constraint on the system operating conditions are investigated, demonstrating an uncertainty level-dependent cost increment for stability maintenance under the distributional robust formulation.

Future work involves the investigation of other distributionally robust approaches and studying the distributionally robust learning formulation as discussed in Section~\ref{sec:5}, where an efficient approach to deal with the nonlinear relationship with the uncertain parameters in the weighted and constrained regression problem will be investigated.

% References section
\bibliographystyle{IEEEtran}
\bibliography{bibliography}
\end{document}

\begin{varwidth}{\linewidth}

\begin{tikzpicture}
\begin{axis}[
    scaled ticks=false,
    tick label style={/pgf/number format/fixed},
    colormap name=viridis,
    width=7.25cm,
    height=4cm,
    %trim axis left,
    %trim axis right,
    %xlabel style={yshift=0.75em},
    %ylabel style={yshift=-1.5em},
    xlabel={$\mathrm{Vulnerable\,Load\,[\%]}$},
    ylabel={$\mathrm{Cost\, Increment\,[k\pounds/h]}$},
    xmin=0, xmax=50,
    ymin=0, ymax=12,
    % xtick={1.2,1.6,2.0,2.4,2.8},
    % ytick={0,1,2,3},
    % yticklabel style={/pgf/number format/.cd,fixed,precision=3},
    %legend pos=north east,
    xmajorgrids=true,
    ymajorgrids=true,
    legend style={at={(axis cs:0.6, 11.65)},anchor=north west,nodes={scale=0.75, transform shape}, legend columns=2},
    legend cell align={left},
    grid style=dashed,
]
\footnotesize
\addplot[
    % smooth,
    thick,
    color=pRed,
    %fill=black, 
    %fill opacity=0.15
    ]
    table {data/Cost_Wind/data2.txt};        
    \addlegendentry{\footnotesize }
\addplot[
    % smooth,
    thick,
    color=pRed,
    dashed,
    %fill=black, 
    %fill opacity=0.15
    ]
    table {data/SI/data0.txt};        
    \addlegendentry{\footnotesize  $3\,\mathrm{GW}$}
\addplot[
    % smooth,
    thick,
    color=pYellow,
    %fill=black, 
    %fill opacity=0.15
    ]
    table {data/Cost_Wind/data3.txt};        
    \addlegendentry{\footnotesize }
\addplot[
    % smooth,
    thick,
    dashed,
    color=pYellow,
    %fill=black, 
    %fill opacity=0.15
    ]
    table {data/SI/data1.txt};     
    \addlegendentry{\footnotesize  $5\,\mathrm{GW}$}
\addplot[
    smooth,
    thick,
    color=pBlue,
    %fill=black, 
    %fill opacity=0.15
    ]
    table {data/Cost_Wind/data4.txt};        
    \addlegendentry{\footnotesize  }
\addplot[
    smooth,
    thick,
    dashed,
    color=pBlue,
    %fill=black, 
    %fill opacity=0.15
    ]
    table {data//SI/data2.txt};        
    \addlegendentry{\footnotesize  $10\,\mathrm{GW}$}
\end{axis}

\end{tikzpicture} 

\end{varwidth}